\documentclass[aps, prl, twocolumn, showpacs, superscriptaddress]{revtex4}
 
\usepackage[dvips]{graphicx}
\usepackage{latexsym}

\begin{document}

\title{Current Flow in Random Resistor Networks: \\
The Role of Percolation in Weak and Strong Disorder}

\author{Zhenhua Wu}
\author{Eduardo L\'{o}pez}
\affiliation{Center for Polymer Studies, Boston University, Boston,
  Massachusetts 02215, USA} 
\author{Sergey V. Buldyrev}
\affiliation{Center for Polymer Studies, Boston University, Boston,
  Massachusetts 02215, USA} 
\affiliation{Department of Physics, Yeshiva University, 500 West 185th
  Street, New York, NY 10033, USA}
\author{Lidia A. Braunstein}
\affiliation{Center for Polymer Studies, Boston University, Boston,
  Massachusetts 02215, USA}
\affiliation{Departamento de F\'{\i}sica, Facultad de Ciencias Exactas y
  Naturales, Universidad Nacional de Mar del Plata, Funes 3350, 7600 Mar del
  Plata, Argentina}
\author{Shlomo Havlin}
\affiliation{Minerva Center of Department of Physics, Bar-Ilan
  University, Ramat Gan, Israel}
\author{H. Eugene Stanley}
\affiliation{Center for Polymer Studies, Boston University, Boston,
  Massachusetts 02215, USA}

%\date{wlbbhs.tex ~~ Last revised: 10:00PM, 4 October 2004}

\begin{abstract}

We study the current flow paths between two edges in a random resistor
network on a $L\times L$ square lattice. Each resistor has resistance
$e^{ax}$, where $x$ is a uniformly-distributed random variable and $a$
controls the broadness of the distribution.  We find (a) the scaled variable
$u\equiv L/a^\nu$, where $\nu$ is the percolation connectedness exponent,
fully determines the distribution of the current path length $\ell$ for all
values of $u$. For $u\gg 1$, the behavior corresponds to the weak disorder
limit and $\ell$ scales as $\ell\sim L$, while for $u\ll 1$, the behavior
corresponds to the strong disorder limit with $\ell\sim
L^{d_{\mbox{\scriptsize opt}}}$, where $d_{\mbox{\scriptsize opt}} =
1.22\pm0.01$ is the optimal path exponent.  (b) In the weak disorder regime,
there is a length scale $\xi\sim a^\nu$, below which strong disorder and
critical percolation characterize the current path.

\end{abstract}

\pacs{64.60.-i, 05.50.+q, 71.30.+h, 73.23.-b, 05.45.Df}
\keywords{Strong disorder, Random resistor network, Optimal path, Transport}

\maketitle

Transport in disordered media is a classic problem in statistical physics
which attracts much attention due to its broad range of
applications. Examples include flow through porous material, oil production,
and conductivity of semiconducting materials or metal-insulator mixtures
\cite{Ambegaokar, Strelniker, Cohen:Ni, Kirkpatrick, Bernasconi, Park,
Berman, Ball, Tyc, Meir}. These problems have been studied using a random
resistor network model with bonds that have a resistance chosen from a
probability distribution mimicking the nature of the physical problem under
consideration.  Among the different classes of disorder distributions used,
the most common is {\it percolation\/} disorder, in which the resistance of a
bond is either 1 or $\infty$~\cite{Bunde}. {\it Gaussian\/} distributions and
{\it power law\/} distributions have also been studied
extensively~\cite{Halpin, Hansen}.

Here, we study a random resistor network with {\it exponential\/}
disorder~\cite{Stauffer}. We consider the two opposite edges of a $L\times L$
square lattice as source $A$ and sink $B$. Each bond connecting adjacent
nodes $i$ and $j$ corresponds to one resistor, whose resistance $r_{ij}$ is
given by \cite{Ambegaokar, Strelniker, Bernasconi, Meir,
fn_exponential_disorder}
\begin{equation}
  r_{ij} = e^{a x_{ij}},
\label{disorder}
\end{equation}
where $a$ controls the disorder strength and $x_{ij}$ is a random number
taken from a uniform distribution $x_{ij}\in [0,1]$. Recent experiments show
that for quenched condensed granular Ni thin films, the conductivity is well
described by exponential disorder with large
$a$~\cite{Strelniker}. Exponential disorder enables us to understand the
magnetoresistance phenomenon that out of $10^9$ grains, only a few govern the
electric conductivity~\cite{Cohen:Ni}. Optimal paths in networks have also
been studied with exponential disorder, where the optimal path is the path
between two sites that minimizes the total weight $\sum_{\mbox{\scriptsize
path}}e^{a x_{ij}}$ \cite{Lidia, Dobrin, Porto, Barabasi, Cieplak}, where the
sum is over the bonds ($ij$) along the path. The length of the optimal path
$\ell_{\mbox{\scriptsize opt}}$ has been shown to scale with the system size
as $L^{d_{\mbox{\scriptsize opt}}}$ for the strong disorder limit ($a \to
\infty$)~\cite{fn_opt_exponent}, where a single bond dominates the optimal
path (and conductance as we see below). The strong disorder limit only has
been related to critical percolation~\cite{Ambegaokar, Strelniker,Bernasconi,
Dobrin}.

Here we show that for exponential disorder, the flow paths for all values of
$a$ are controlled by critical percolation and by the scaling properties of
the optimal path in the strong disorder limit. Indeed, the resistance of each
path is equal to the sum of its resistances. When $a \to \infty$ the
resistance of each path is dominated by the largest resistance on this path
$\exp(a x_{\rm max})$. Almost all currents must go along the path which
minimizes $x_{\rm max}$. We denote this min-max value of disorder as $x_1
\equiv \min_{\mbox{\scriptsize all paths}} x_{\rm max}$.  Among all the paths
which go through the bond with $x_1$, the maximum-current goes along the path
which minimizes the second largest value of disorder $x'_{\rm max}$, and so
on. Thus the algorithm of selecting the path with the maximum-current is
equivalent to selecting the optimal path in the strong disorder limit
(ultrametric algorithm~\cite{Cieplak}). As $a \to \infty$ the maximum-current
path coincides with the optimal path in the strong disorder limit.  On the
other hand, since all values $x_{ij}$ on the maximum-current path are below
$x_1$, this path must belong to the percolation backbone with concentration
$p$ equal to the fraction of bonds whose $x_{ij}<x_1$~\cite{fn_stauffer}. The
value of $p$ at which percolation between two edges of the system does occur
has a narrow distribution with a mean of $p=p_c$ and a standard deviation
that scales as $\sim L^{-1/\nu}$~\cite{Coniglio}, where $p_c$ is the critical
percolation threshold, $L$ is the linear system size, and $\nu$ is the
connectedness length exponent.  Thus the value of $x_1$ also must have a
narrow distribution of width $\sim L^{-1/\nu}$.

Next we estimate the value $a$ at which the maximum-current path starts to
bifurcate. Consider the paths which do not pass through bond $x_1$ as if this
bond has been cut \cite{Strelniker}. The maximum-current will then pass
through bond $x_2>x_1$, which is characterized by the same narrow
distribution. Hence $(x_2-x_1)$ is of the order of $L^{-1/\nu}$. These paths
become competitive with the true optimal path if its resistance $\exp(a x_2)$
becomes of the same order as $\exp(a x_1)$ or if $a(x_2-x_1) \approx
aL^{-1/\nu}\approx 1$. This condition determines the crossover from weak to
strong disorder. If $L \ll a^\nu$, the disorder is strong and the
maximum-current path does not bifurcate. If $L \gg a^\nu$ the disorder is
weak and the maximum-current path can bifurcate. Moreover, the value $\xi
\sim a^\nu$ determines the connectedness length below which the disorder is
strong and the maximum-current path is determined by the unique optimal path
and above which the maximum-current path bifurcates.

\begin{figure}
\includegraphics[width=0.235\textwidth]{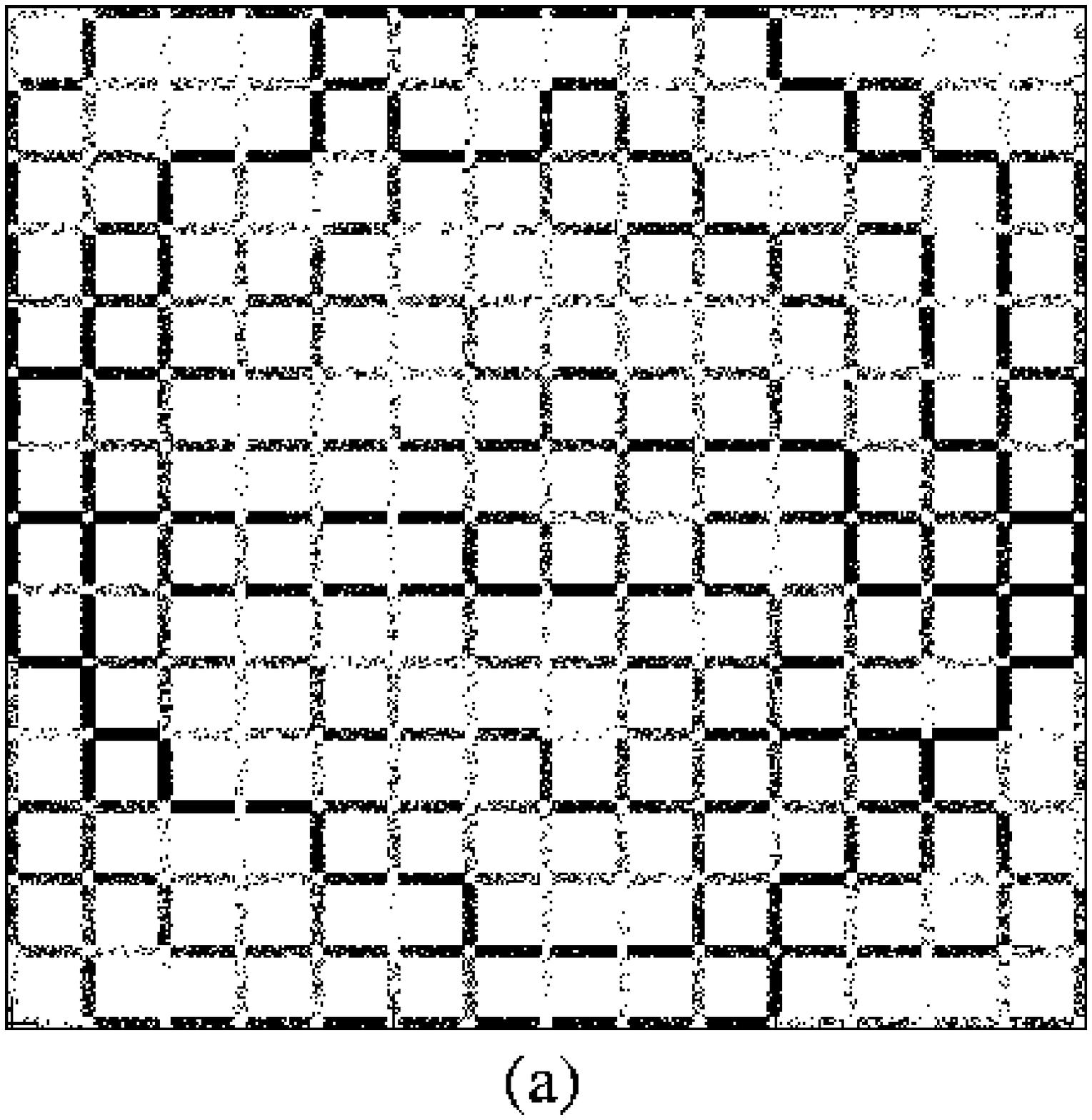}
\includegraphics[width=0.235\textwidth]{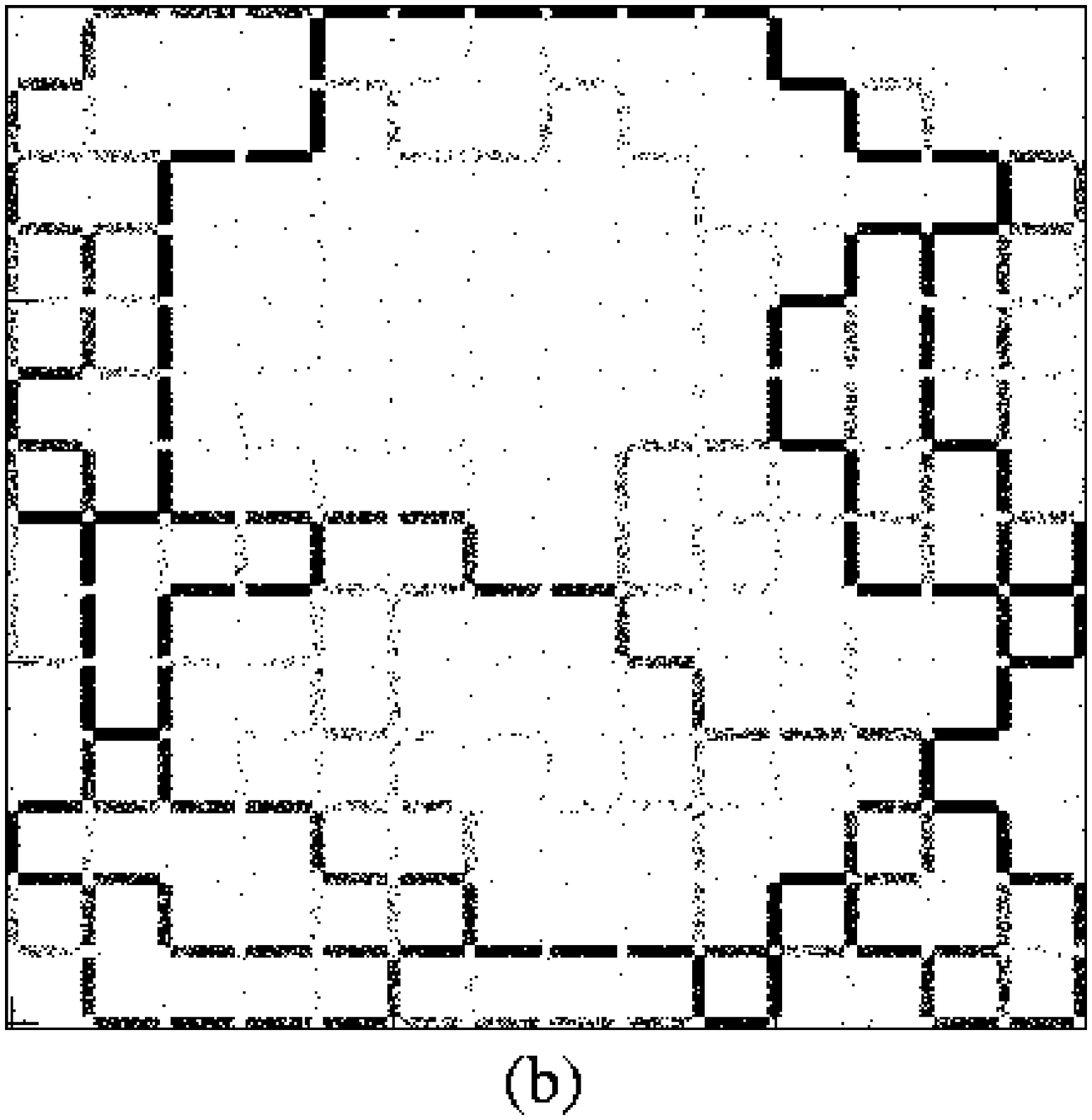}
\includegraphics[width=0.235\textwidth]{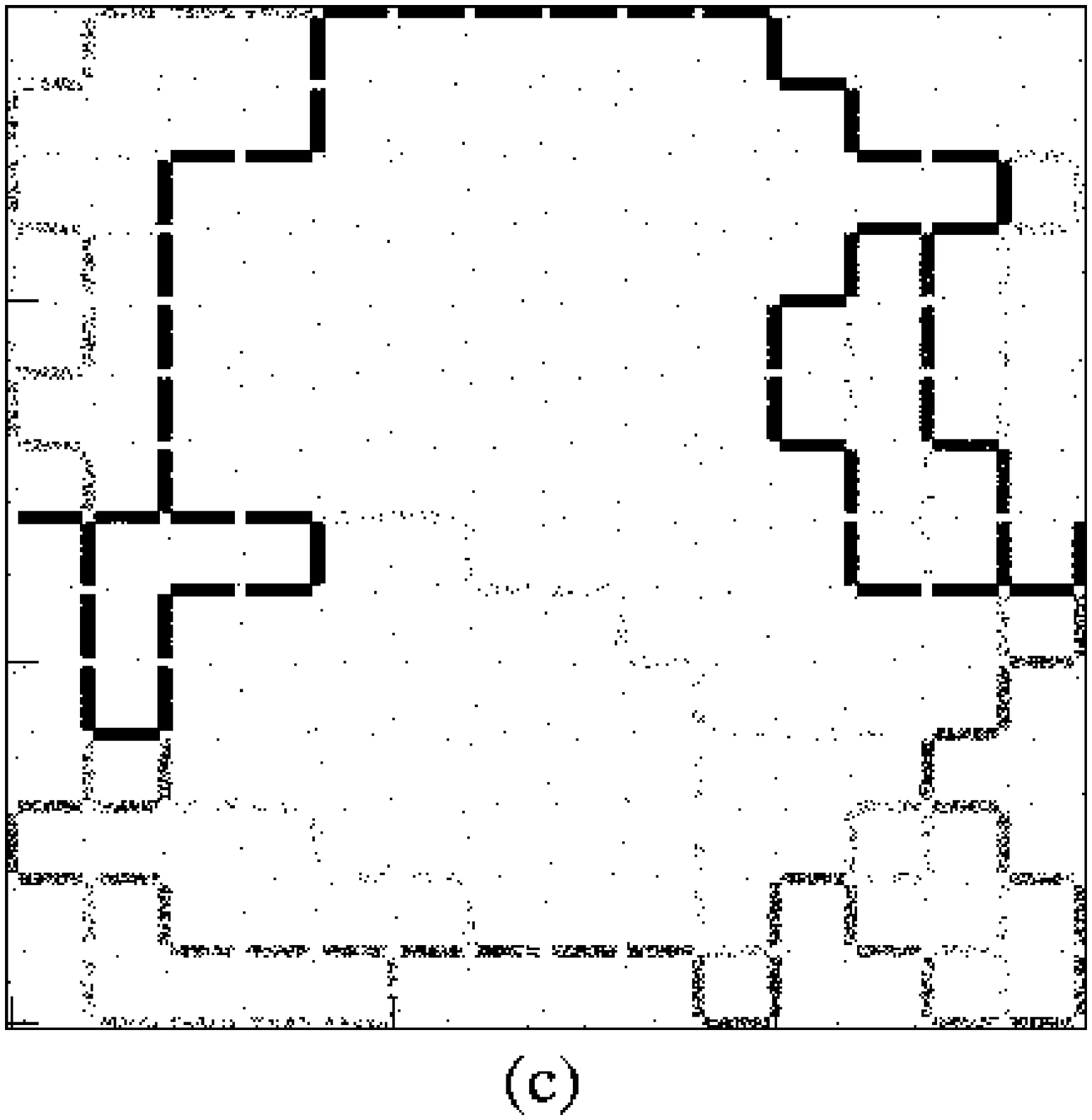}
\includegraphics[width=0.235\textwidth]{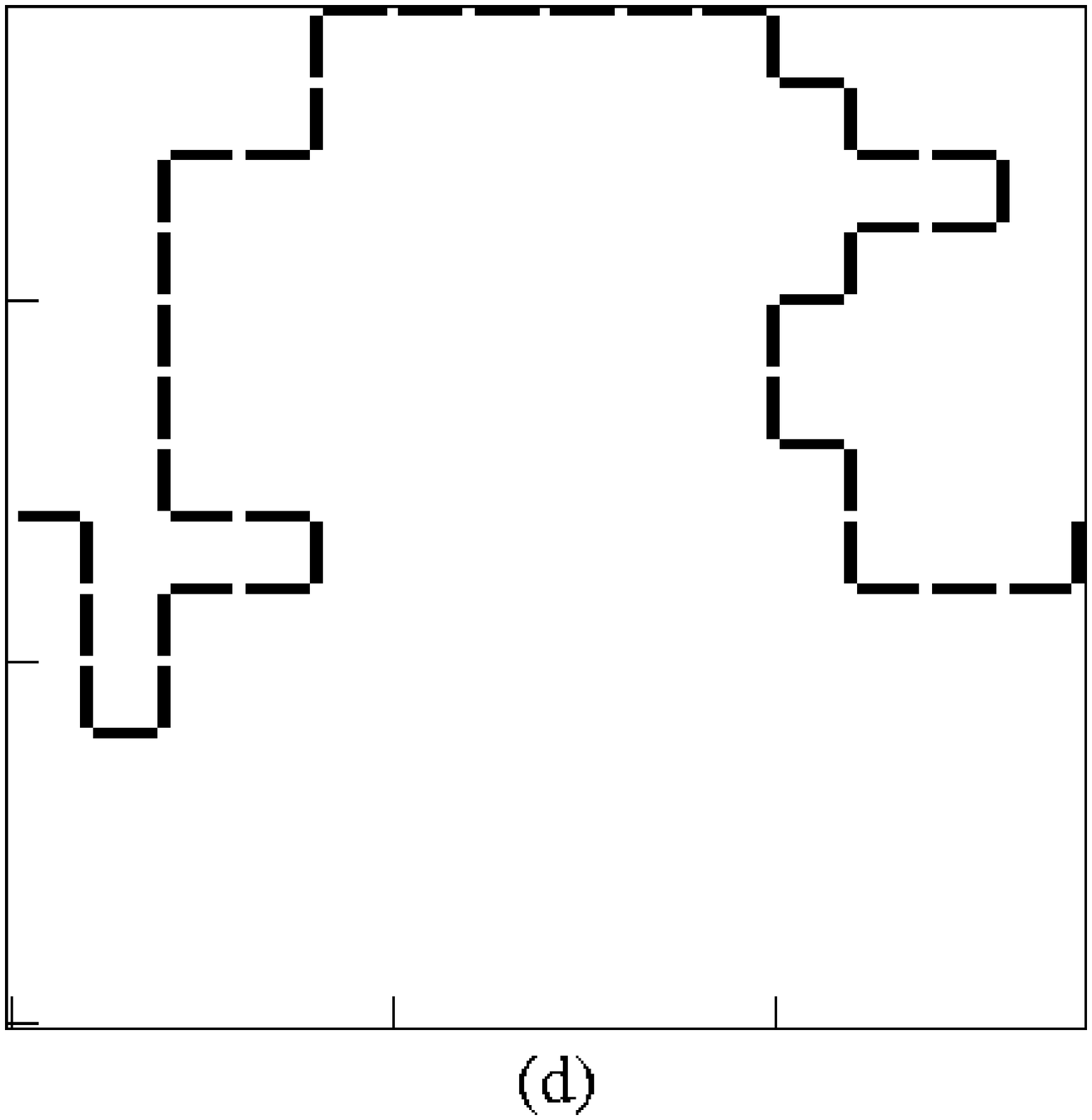}
\caption{\label{current_map}Current maps for the same configuration of
disorder on a 15x15 square lattice, with different values of $a$: (a)
$a=5$, (b) $a=20$, and (c) $a=45$. Each bond represents one
resistor. The dot density of each bond increases when the bond current
increases. The source has coordinate (0,7) and the sink (14,7). (d) The
corresponding optimal path for the same configuration of disorder and
for $a=45$. The similarity between (c) and (d) suggests a relation
between current flow paths for large $a$ and the optimal path.}
\end{figure}

To confirm these analytical predictions, we study the problem
numerically. Define the electric potential at node $i$ of the lattice as
$V_i$, and set the potentials at source and sink as $V_A=1$ and $V_B=0$, we
numerically solve the set of Kirchhoff equations for all
$V_i$~\cite{Kirchhoff_equation}. We begin by building an intuitive
understanding of the effect of changing the strength of disorder on current
flow. Figures~\ref{current_map}(a), (b), and (c) show, for different values
of $a$, the magnitudes of the bond currents represented by the density of
dots on each bond. We see that the set of bonds carrying most of the current
decreases as $a$ increases, so that only a few current paths dominate. This
confirms earlier findings that for large $a$, one or very few paths dominate
the current flow~\cite{Ambegaokar, Strelniker, Bernasconi, Berman}. In
Fig.~\ref{current_map}(d), we plot the optimal path for the same disorder
realization. The similarity between the path of the current carrying bonds in
Fig.~\ref{current_map}(c) and the optimal path in Fig.~\ref{current_map}(d)
exhibits how these two quantities are related in the strong disorder limit
and supports the argument above that the maximum-current path coincides with
the optimal path.

\begin{figure}
  \includegraphics[width=0.18\textwidth, angle=-90]{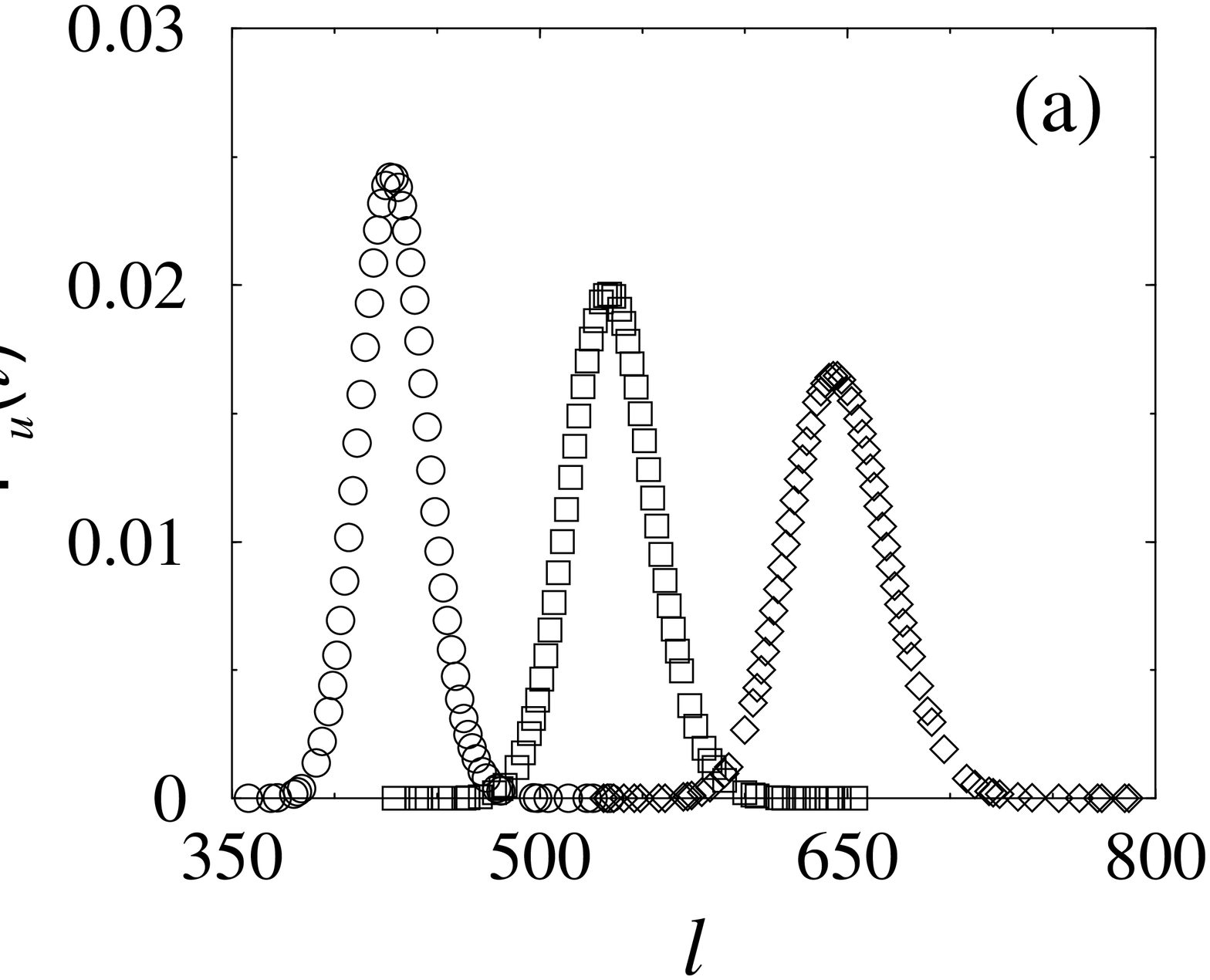}
  \includegraphics[width=0.18\textwidth, angle=-90]{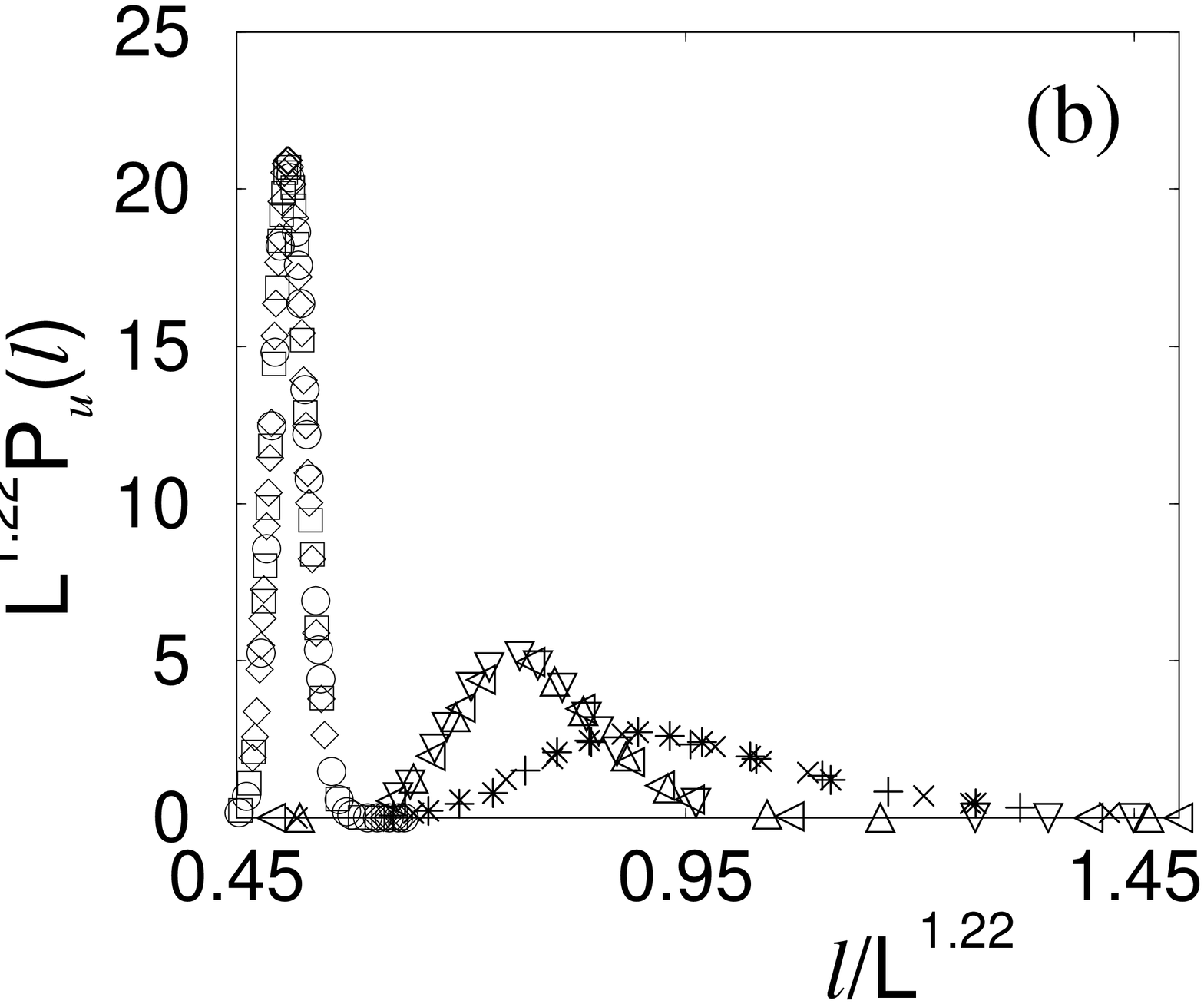}
  \caption{\label{distr_scaling}(a) Plot of $P(\ell \mid L, u)$ versus $\ell$
for square lattices with fixed $u\equiv L/a^\nu=10$ and different values of
$L$. (b) Plot of $P(\ell \mid L, u) L^{1.22}$ versus $\ell/L^{1.22}$. Three
families of curves are shown and each family has the same ratio $u \equiv
L/a^\nu$: $u = 10$ [$L=250(\circ)$, $L=300(\Box)$, $L=350(\diamond)$]; $u =
1.26$ [$L=30(\triangle)$, $L=40(\triangleleft)$, $L=60(\bigtriangledown)$];
$u = 0.25$ [$L=15(+)$, $L=20(\times)$, $L=27(*)$]. The distribution curves
with the same ratio of $u$ collapse both in weak disorder such as $u = 10$
and in strong disorder such as $u = 0.25$, as well as in the intermediate
regime $u = 1.26$ for $a>10$ and $L>15$. We compute all the data with 1000
realizations of disorder and $10^5$ tracers for each realization.}
\end{figure}

Figure~\ref{current_map} illustrates that the paths used by the current are
intimately related to the disorder of the system. Therefore, we study the
ensemble of current paths on the lattice by performing tracer dynamics with
the particle launching algorithm~\cite{Eduardo}. For a given realization, all
bond currents are determined by Kirchhoff equations and then tracers are
injected into node $A$ and extracted at node $B$. At a given node, the tracer
follows the bond from node $i$ to $j$ with probability
\begin{equation}
\omega_{ij} = {J_{ij} \over \sum_j J_{ij}},
\end{equation}
where $j$ runs over all the neighbor bonds of node $i$,
$J_{ij} = I_{ij}$ if $I_{ij}\geq 0$, and $J_{ij}=0$ if $I_{ij} < 0$, so that
only ``out'' currents are taken into account.

To understand the behavior of the current flow in the presence of disorder in
all ranges of disorder, we calculate the length distribution of all tracer
paths, $P(\ell\mid L,a)$, from $A$ to $B$ for a system of linear size $L$ and
disorder strength $a$. We first fix $u \equiv L/a^\nu$ and calculate the
distribution $P_u(\ell)\equiv P(\ell \mid L,a)$ for different system sizes
$L$ and the corresponding values of $a = (L/u)^{1/\nu}$. We obtain weak
disorder when $u \gg 1$ and strong disorder when $u \ll 1$, as found for the
optimal path in networks \cite{Sameet} and as shown below for current
flow. Moreover, we find that $u$ is the only parameter that characterizes the
disorder and thus determines $P_u(\ell)$.

In Fig.~\ref{distr_scaling}(a) we show three normalized distributions
$P_u(\ell)$ with $u = 10$ (weak disorder), which collapses to a single
curve as shown in Fig.~\ref{distr_scaling}(b).
Figure~\ref{distr_scaling}(b) also shows two other peaked curves with
$u=1.26$ (close to the crossover) and $u=0.25$ (strong disorder). Each
curve shows the collapse of three distributions with different system
sizes $L$ but the same value of $u$. This collapse implies that $P(\ell
\mid L, a)$ is controlled by a single parameter~$u$
\begin{equation}
  P(\ell\mid L, a)\sim \frac{1}{L^{d_{\mbox{\scriptsize opt}}}}
  f_u\left(\frac{\ell}{L^{d_{\mbox{\scriptsize opt}}}}\right)
  \label{distr_scale_func}.
\end{equation}
We confirmed this scaling numerically for values of $u$ between $u = 10$
(weak disorder) and $u = 0.25$ (strong disorder) for $a>10$ and
$L>15$~\cite{fn_weak_deviation}.

\begin{figure}
  \begin{center}
    \includegraphics[width=0.45\textwidth]{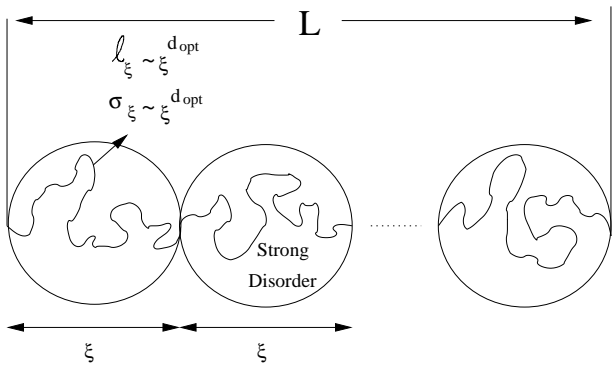}
    \caption{\label{weak_scheme} Schematic illustration of the flow path
    inside and outside the critical regimes in the weak disorder case, $\xi =
    a^\nu \ll L$. The parameter $u = L/a^\nu$ determines the number of such
    units (of size $\xi$) in a linear size $L$. While the total length of the
    flow path is linear with $L$, for distance $y < \xi$ (inside the critical
    regime), we expect $\ell \sim y^{d_{\mbox{\scriptsize opt}}}$.}
  \end{center}
\end{figure}

To understand why $\nu$ and $d_{\mbox{\scriptsize opt}}$ play an important
role in determining the length of the current flow path in weak disorder as
well as in strong disorder, we suggest the following theoretical argument. In
the weak disorder regime, there is a characteristic length $\xi \sim a^\nu$
below which strong disorder exists and critical percolation plays a crucial
role~\cite{fn_case_strong}. We thus expect that for length scales up to $\xi$
the tracers travel on strong disorder path segments with a typical length of
$\ell_{\xi} \sim \xi^{d_{\mbox{\scriptsize opt}}}$, and a tracer length
deviation of $\sigma_{\xi} \sim \xi^{d_{\mbox{\scriptsize opt}}}$
(illustrated in Fig.~\ref{weak_scheme}). For a system of linear size $L$ in
weak disorder, the ratio of the system size to the connectedness length $u
\sim L/\xi$ roughly indicates the number of independent strong disorder
tracer path segments within a complete tracer path from source to sink. The
total length is obtained by multiplying $u$ by the length of a segment,
$\xi^{d_{\mbox{\scriptsize opt}}}$. Defining $\ell^*$ as the maximum of
$P(\ell \mid L, a)$, we thus predict that in the weak disorder $\ell^*$ is
\begin{equation}
  \ell^* \sim \overline{\ell} \sim u\xi^{d_{\mbox{\scriptsize opt}}} =
  L^{d_{\mbox{\scriptsize opt}}} u^{1 - d_{\mbox{\scriptsize opt}}}
  \label{clt_mean}\;,
\end{equation}
where $\overline{\ell}$ is the mean average path length of the
tracers~\cite{lstar_explain, central_limit}.
Thus for all values of $u$, $\ell^*$ can be written in a unified form
\begin{equation}
  \ell^* \sim L^{d_{\mbox{\scriptsize opt}}} g_{\ell}(u),
  \label{scale_ansatz_ell}
\end{equation}
where $g_{\ell}(u)$ is a scaling function that satisfies (from
Eq.~(\ref{clt_mean}))
\begin{equation}
  g_{\ell}(u) \sim \left\{ \begin{array}{ll}
    u^{1 - d_{\mbox{\scriptsize opt}}}  & u \gg 1 \\
    1 & u \ll 1 .
  \end{array}\right.
  \label{scale_function_ell}
\end{equation}
The arguments leading to Eq.~(\ref{clt_mean}) for weak disorder, also
imply that the standard deviation $\sigma$ of $\ell$ scales as
\begin{equation}
  \sigma \sim \sqrt{u} \xi^{d_{\mbox{\scriptsize opt}}} =
	L^{d_{\mbox{\scriptsize opt}}} u^{1/2 - d_{\mbox{\scriptsize opt}}},
  \label{clt_sigma}
\end{equation}
and for all values of $u$
\begin{equation}
  \sigma \sim L^{d_{\mbox{\scriptsize opt}}} g_{\sigma}(u),
  \label{scale_ansatz_sigma}
\end{equation}
with
\begin{equation}
  g_{\sigma}(u) \sim \left\{ \begin{array}{ll}
    u^{1/2 - d_{\mbox{\scriptsize opt}}}  & u \gg 1 \\
    1 & u \ll 1 .
  \end{array}\right.
  \label{scale_function_sigma}
\end{equation}

\begin{figure}
\begin{center}
  \includegraphics[width=0.18\textwidth, angle=-90]{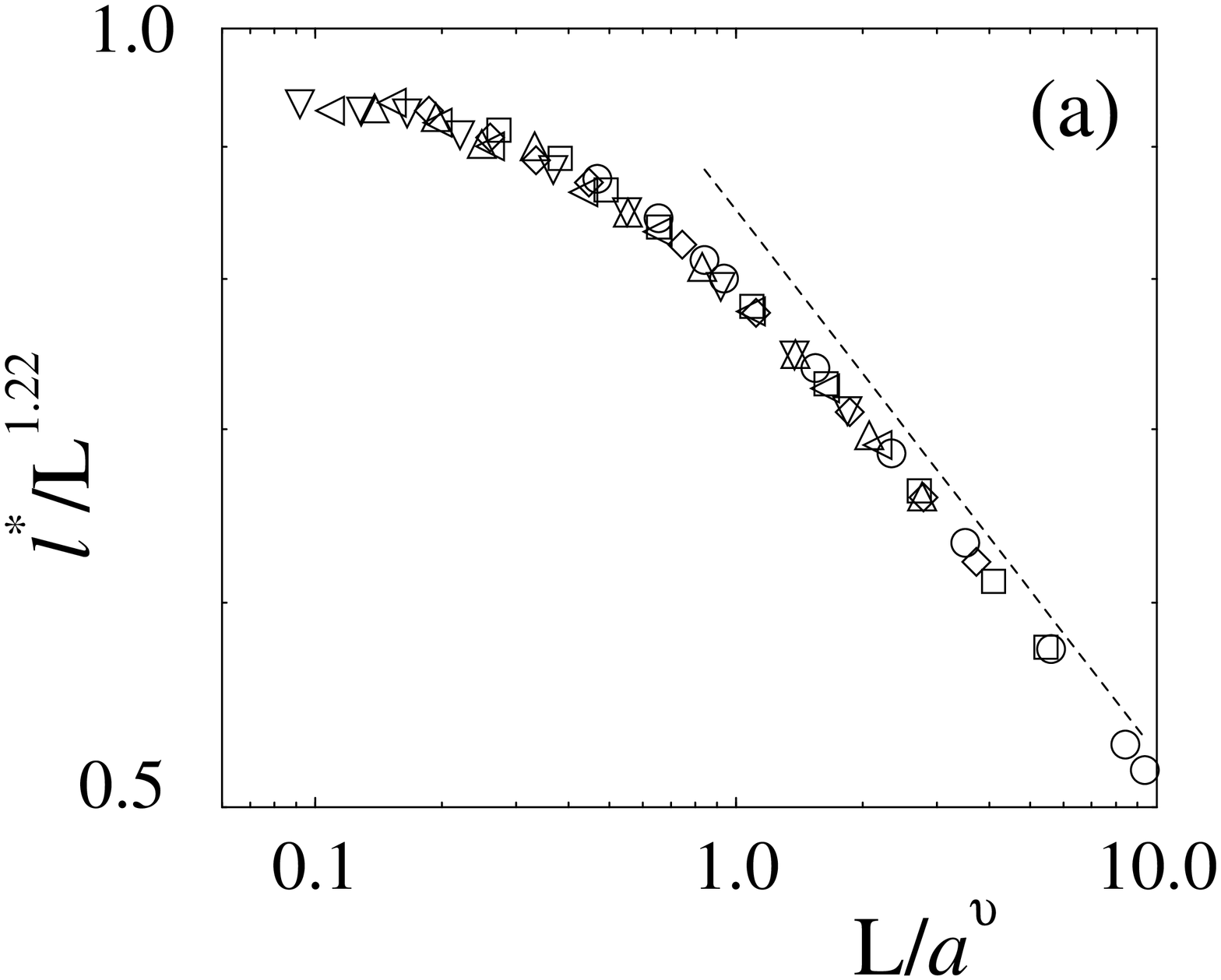}
  \includegraphics[width=0.18\textwidth, angle=-90]{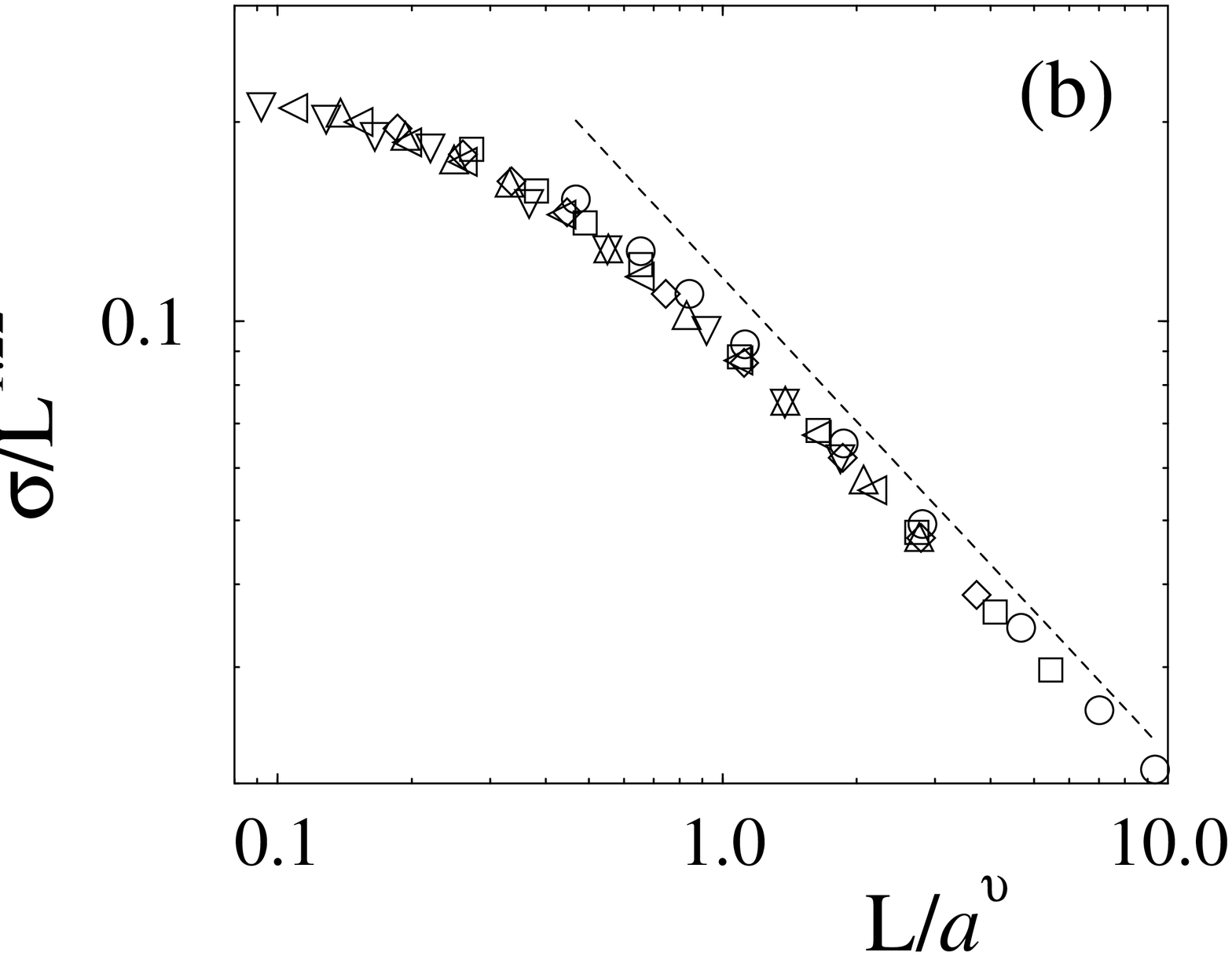}
  \caption{\label{lstar_width}(a) Log-log plot of $\ell^*/L^{1.22}$ versus
    $L/a^{\nu}$ for different values of $a$: 10($\circ$), 15($\Box$),
    20($\diamond$), 25($\triangle$), 30($\triangleleft$),
    34($\bigtriangledown$) and $L$ changes from $20$ to $200$. The slope of
    the dashed line is -0.21, in agreement with Eqs.~(\ref{scale_ansatz_ell})
    and (\ref{scale_function_ell}). (b) The same as (a) but for $\sigma$. The
    slope of the dashed line is -0.72, in agreement with
    Eqs.~(\ref{scale_ansatz_sigma}) and (\ref{scale_function_sigma}).}
\end{center}
\end{figure}

To test Eq.~(\ref{scale_ansatz_ell}) we plot $\ell^*/L^{d_{\mbox{\scriptsize
opt}}}$ as a function of $u$ in Fig.~\ref{lstar_width}(a). We find that the
best scaling is obtained for $d_{\mbox{\scriptsize opt}} = 1.22$, the
predicted value.  When $L \gg a^{\nu}$ $(u \gg 1)$, $g_{\ell}(u)$ is
asymptotically a power law function with an exponent $-0.21 \pm 0.02$, which
is within the error of the predicted value $1 - d_{\mbox{\scriptsize opt}} =
-0.22\pm0.01$ (from Eq.~(\ref{scale_function_ell})). Similarly, in
Fig.~\ref{lstar_width}(b), we plot $\sigma / L^{d_{\mbox{\scriptsize opt}}}$
as a function of $u = L/a^\nu$ and find that $g_{\sigma}$ is asymptotically a
power law with an exponent $-0.72\pm 0.02$ as predicted in
Eq.~(\ref{scale_function_sigma}). All these results strongly support our
picture of critical percolation regimes of size $\xi\sim a^\nu$.

Equations~(\ref{scale_ansatz_ell}) and~(\ref{scale_function_ell}) state that
tracer path length scales with system size $L$ in the same way as the optimal
path length for all values of $u$. For $u \ll 1$, $\ell^* \sim
L^{d_{\mbox{\scriptsize opt}}}$ and the path is a fractal with the same
exponent $d_{\mbox{\scriptsize opt}}$ as for the optimal path length
$\ell_{\mbox{\scriptsize opt}}$. In weak disorder ($u \gg 1$), we obtain
$\ell^* \sim L$ as we do for self-affine structures~\cite{Bunde}. This
is consistent with the interesting possibility that they belong to the same
universality class. As $u \ll 1$, current flows only along the optimal path,
which explains the existence of the bottleneck at the percolation threshold
$p_c$~\cite{Ambegaokar, Strelniker, Berman, Bernasconi}.

Our results also explain the simulation results of Ref.~\cite{Strelniker} for
the scaled plot $\log(R_{\mbox{\scriptsize cut}}/R)$ as a function of
$a/L^{1/1.3}$, where $R$ is the equivalent resistance of the 2D random
resistor lattice and $R_{\mbox{\scriptsize cut}}$ is the equivalent
resistance of the system after cutting the bond with the maximal local
current. Before cutting this bond, the equivalent resistance in the strong
disorder limit is dominated by the maximal resistance along the optimal path
$R \sim e^{a p_c}$~\cite{Ambegaokar, Bernasconi, Berman}. After cutting this
bond, the current will reorganize to follow a new optimal path on which the
dominant resistance is $R \sim e^{a p}$, making the ratio
$R_{\mbox{\scriptsize cut}}/R \sim e^{ a (p-p_c)} = e^{a \delta p}$. Using
the relation $\delta p\sim \xi^{-1/\nu} \sim L^{-1/\nu}$~\cite{Coniglio}, we
find that 
\begin{equation}
  {R_{\mbox{\scriptsize cut}} \over R} \sim e^{a / L^{1/\nu}}\;.
\end{equation}
This result also analytically supports our assumption that the ratio $L/a^\nu$
characterizes the disorder and determines the properties of current flow.

In summary, we find that the tracer path length $\ell$ in flow in the
presence of exponential disorder behaves similarly to the optimal path length
$\ell_{\mbox{\scriptsize opt}}$, and even has the same scaling exponents
($d_{\mbox{\scriptsize opt}}$ for $u \ll 1$ and one for $u \gg 1$). Moreover,
we also find that when the disorder is weak and $\ell \sim L$, there is a
connectedness length $\xi\sim a^\nu$, where strong disorder and critical
percolation exist for regimes smaller than $\xi$. As a result, the
probability distribution of $\ell$ is determined by the ratio $L/a^\nu$,
which is the number of units of size $\xi\sim a^\nu$ in a linear size $L$.

We thank S. Sreenivasan and G. Paul for useful discussions, ONR,
ONR-Global, and the Israel Science Foundation for financial support.


\begin{thebibliography}{50}
\bibitem{fn_exponential_disorder} This class corresponds to a
distribution of resistances, which behaves as $1/r$ below a cutoff value
$r \leq e^a$, and zero for $r > e^a$.
\bibitem{fn_opt_exponent} The values of the exponent
$d_{\mbox{\scriptsize opt}}$ are: $d_{\mbox{\scriptsize opt}} = 1.22
(d=2)$ and $1.43 (d=3)$ see ref~\cite{Cieplak}. References~\cite{Porto}
and~\cite{Sameet} show that when $L/a^\nu \ll 1$ these results hold
(strong disorder) while for $L/a^\nu \gg 1$, $\ell_{\mbox{\scriptsize
opt}} \sim L$ which is the weak disorder regime.
\bibitem{fn_stauffer} In Ref.~\cite{Stauffer} similar considerations are used
to calculate the average conductance of this model, but not the properties of
flow paths. The latter have been mostly ignored in the literature. However,
recent studies regarding the optimal path in strong disorder and our own new
insight allow us to explain the properties of the flow paths.
\bibitem{Kirchhoff_equation} The current between nodes $i$ and $j$ is $I_{ij}
= (V_i - V_j) / r_{ij}$. The Kirchhoff equations become $\sum_j{I_{ij}} =
I(\delta_{i,A} - \delta_{i,B})$, where $\delta$ is the Kronecker delta and
$I$ is the total current from $A$ to $B$.
\bibitem{fn_case_strong} In the strong disorder $\xi = a^\nu > L$ and
  therefore the whole system is in strong disorder.
\bibitem{lstar_explain} Because the whole distribution scales the same way
  according to Eq.~(\ref{distr_scale_func}), the mean value $\overline{\ell}$
  and the most probable value $\ell^*$ have the same scaling.
\end{thebibliography}
\end{document}